\documentclass[12pt]{iopart}

\usepackage{graphicx}
\begin{document}

\title[The 600 K T Dwarfs]{The Physical Properties of Four
$\sim$600 K T Dwarfs}

\author{S K Leggett$^1$, Michael C Cushing$^2$, D Saumon$^3$,  M S Marley$^4$, 
T L Roellig$^4$, S J Warren$^5$, Ben Burningham$^6$, H R A Jones$^6$, J D Kirkpatrick$^7$,
N Lodieu$^8$,
P W Lucas$^6$,  A K Mainzer$^9$, E L Martin$^8$, M J McCaughrean$^{10}$, D J Pinfield$^6$,  
G C Sloan$^{11}$,  R L Smart$^{12}$, M Tamura$^{13}$ and J Van Cleve$^{14}$ }

\address{$^1$ Gemini Observatory, 670 N. A ohoku Place, Hilo, HI 96720, USA}
\address{$^2$ Institute for Astronomy, University of Hawaii, 2680 Woodlawn Drive, Honolulu, HI 96822}
\address{$^3$ Los Alamos National Laboratory, P.O. Box 1663, MS F663, Los Alamos, NM 87545}
\address{$^4$ NASA Ames Research Center, Mail Stop 245-3, Moffett Field, CA 94035}
\address{$^5$ Imperial College London, Blackett Laboratory, Prince Consort Road, London SW7 2AZ}
\address{$^6$ Centre for Astrophysics Research, Science and Technology Research Institute, University of Hertfordshire, Hatfield AL10 9AB}
\address{$7$ IPAC, California Institute of Technology, Mail Code 100-22, 770 South Wilson Avenue, Pasadena, CA 91125}
\address{$8$ Instituto de Astrofsica de Canarias, 38200 La Laguna, Spain}
\address{$9$ Jet Propulsion Laboratory, California Institute of Technology, 4800 Oak Grove Drive, Pasadena, CA 91109}
\address{$10$ School of Physics, University of Exeter, Stocker Road, Exeter, Devon EX4 4QL}
\address{$11$ Astronomy Department, Cornell University, Ithaca, NY 14853}
\address{$12$ INAF-Osservatorio Astronomico di Torino, Strada Osservatorio 20, 10025 Pino Torinese, TO, Italy}
\address{$13$ National Astronomical Observatory, Mitaka, Tokyo 181-8588}
\address{$14$ Ball Aerospace and Technologies Corp., Boulder, CO 80301}

\ead{sleggett@gemini.edu}

\begin{abstract} We present $Spitzer$ 7.6 -- 14.5 $\mu m$ spectra of ULAS 
J003402.77$-$005206.7 and ULAS J133553.45$+$113005.2, two T9 dwarfs with the latest 
spectral types currently known.  We fit synthetic spectra and photometry to the near- 
through mid-infrared energy distributions of these dwarfs and that of the T8 dwarf 2MASS 
J09393548$-$2448279. We also analyse near-infrared data for another T9, CFBD 
J005910.82$-$011401.3. We find that the ratio of the mid- to near-infrared fluxes is very 
sensitive to effective temperature at these low temperatures, and that the 2.2~$\mu$m and 
4.5~$\mu$m fluxes are sensitive to metallicity and gravity; increasing gravity has a 
similar effect to decreasing metallicity, and vice versa, 
and there is a degeneracy between these parameters. 
The 4.5~$\mu$m and 10~$\mu$m fluxes are also sensitive to vertical transport of 
gas through the atmosphere, which we find to be significant for these dwarfs. The full 
near- through mid-infrared spectral energy distribution allows us to constrain the 
effective temperature (K) / gravity (ms$^{-2}$) / metallicity ([m/H] dex)\ of ULAS 
J0034$-$00 and ULAS J1335$+$11 to 550--600/100--300/0.0--0.3 and 
500--550/100--300/0.0--0.3, 
respectively.  These fits imply low masses and young ages for the dwarfs of 5 -- 20 
M$_{Jupiter}$ and 0.1 -- 2 Gyr. The fits to 2MASS J0939$-$24 are in good agreement with the 
measured distance, the observational data, and the earlier T8 near-infrared spectral type 
if it is a slightly metal-poor 4 -- 10 Gyr old system consisting of a 500~K and 700~K, 
$\sim$25 M$_{Jupiter}$ and $\sim$40 M$_{Jupiter}$, pair, although it is also possible that it
is  an identical pair of 600~K, 30 M$_{Jupiter}$, dwarfs. 
As no mid-infrared data are available 
for CFBD 
J0059$-$01 its properties are less well constrained; nevertheless it appears to be a 550 -- 
600 K dwarf with $g =$ 300 -- 2000 ms$^{-2}$ and [m/H] $=$ 0 -- 0.3 dex.  These properties 
correspond to mass and age ranges of 10 -- 50 M$_{Jupiter}$ and 0.5 -- 10 Gyr for this dwarf.

\end{abstract}

\pacs{97.10.Ex, 97.20.Vs}


\section{Introduction}

The low-temperature field brown dwarfs  are of interest for many reasons.
They are important for studies of 
star formation and the initial mass function at low masses, and of 
the chemistry and physics of cool  atmospheres.
Furthermore, they populate the
substellar temperature range and can act as proxies for giant exoplanets
which have similar temperatures and radii, but which 
have less well understood interiors and which
are harder to observe than isolated brown dwarfs.

The far-red  Sloan Digital Sky Survey (SDSS, York et al. 2000), and near-infrared  
Deep Near-Infrared Survey of the Southern Sky (DENIS, Epchtein 1998) and
Two Micron All Sky Survey (2MASS, Skrutskie et al. 2006), 
led to the discovery of the field population of very low mass and low temperature
stars and  brown dwarfs -- the L and T dwarfs (e.g. Kirkpatrick et al. 1997,
Martin et al. 1999, Strauss et al. 1999, Leggett et al. 2000). The L dwarfs 
have effective temperatures ($T_{\rm eff}$) of
1400 -- 2300~K and the T dwarfs have $T_{\rm eff} < 1400$~K (Golimowski et al. 2004,
Vrba et al. 2004), with a lower limit not yet defined. 
Burrows, Sudarsky \& Lunine (2003) calculate that several changes will occur near
$T_{\rm eff} = 400$~K which may trigger a new spectral type, including
the formation of H$_2$O clouds, strengthening of NH$_3$ and the increasing importance
of the mid-infrared flux region.

The flux from the
T dwarfs is predominantly emitted in the far-red through mid-infrared, and the 
classification scheme is based on the strength of
the near-infrared H$_2$O and CH$_4$ absorption bands (Burgasser et al. 2006).
At the time of writing, only 14 T dwarfs are known with spectral type later than T7, i.e. with
$T_{\rm eff}$ less than $\sim$900~K (e.g. {\it http://www.DwarfArchives.org}). 
The near-infrared spectra of  three recently 
discovered T9 dwarfs (Warren et al. 2007, Burningham et al.
2008, Delorme et al. 2008a), the latest spectral type currently known, 
hint at an NH$_3$ feature at 1.5~$\mu$m and
suggest that we may be at the brink of defining the next spectral class, 
for which the letter Y has been suggested (Kirkpatrick et al. 1999).

In this paper we present new mid-infrared spectra for two of these T9 dwarfs,
ULAS J003402.77$-$005206.7 and ULAS J133553.45$+$113005.2,
obtained with the Infrared Spectrograph (IRS, Houck et al. 2004) on the
{\it Spitzer Space Telescope} (Werner et al. 2004).  We combine these data with the previously published near-infrared spectra and mid-infrared IRAC photometry 
to generate an almost complete spectral 
energy distribution for the dwarfs, and use atmospheric and evolutionary models  
to constrain their physical properties. We also investigate similar data for
2MASS J09393548$-$2448279, a T8 dwarf recently found to be very cool
(Burgasser et al. 2008), and the near-infrared spectrum for the third T9 dwarf,
CFBD J005910.82$-$011401.3.

\section{The Sample}

Two T9 dwarfs were recently found in the  UK Infrared Telescope (UKIRT) 
Infrared Deep Sky Survey (UKIDSS, Lawrence et al. 2007)
and another in the far-red  Canada-France-Hawaii Telescope's Brown Dwarf Survey 
(CFBDS, Delorme et al. 2008b):
ULAS J003402.77$-$005206.7 (hereafter ULAS J0034$-$00; Warren et al. 2007), 
ULAS J133553.45$+$113005.2 (ULAS J1335$+$11; Burningham et al. 2008), and 
CFBD J005910.82$-$011401.3 (CFBD J0059$-$01; Delorme et al. 2008a).
The two UKIDSS dwarfs were found in the 1$^{\rm st}$ and 3$^{\rm rd}$ data releases 
of the Large Area Survey (LAS) component of UKIDSS, which reaches $J=19.6$. 
The 3$^{\rm rd}$ data release covers 900 deg$^2$ of a planned 4000 deg$^2$ for the LAS. 
The CFBDS reaches $z_{AB}=22.5$, and 350 deg$^2$ have been analysed.
The discovery papers show that the near-infrared spectra, as well as IRAC photometry
for the two UKIDSS dwarfs,
imply a very low $T_{\rm eff}$ of $\sim$600~K for these three dwarfs.
Here we present new data for  the two UKIDSS T9 dwarfs;
no new data are presented for the CFBDS dwarf but we include it in our analysis 
as these three dwarfs appear to be similarly cool. 
We also include in this analysis a dwarf discovered in the 2MASS survey by
Tinney et al. (2005):  2MASS J09393548$-$2448279 (2MASS J0939$-$24). Although
classified as T8 using low-resolution near-infrared spectra and methane imaging, Burgasser et al.
(2008) show that mid-infrared data constrain its temperature to be similar to that
of the T9 dwarfs.

Figure 1 shows the near-infrared spectra of these four dwarfs as well as
the spectrum of the T8 spectral standard 2MASS J04151954$-$0935066 (2MASS J0415$-$09; Burgasser et al. 2002, 2006) which 
has $T_{\rm eff}=$ 725 -- 775~K (Saumon et al. 2007). 
At these low temperatures very little flux remains to be absorbed in the troughs of the 
strong H$_2$O and CH$_4$ bands and  spectral typing based on these fluxes
becomes difficult. Burningham et al. (2008), Delorme et al. (2008a) and 
Warren et al. (2007) show that spectral indices which measure the blue wings
of the $J$ and $H$ flux peaks, ``W$_J$'' at $\approx$1.21~$\mu$m and 
``NH$_3$-$H$'' at $\approx$1.55~$\mu$m, appear to be useful type indicators,
and that the previously defined ``H$_2$O-$H$''  at $\approx$1.50~$\mu$m also retains
some use at these extreme types (see e.g. Figures 7 though 9 of Burningham et al.).
It is likely that there is an  NH$_3$ absorption feature in the blue wing of the $H$ band for the coolest dwarfs, and Delorme et al. (2008a) suggest that these objects may be the prototype of the Y spectral class. However the feature is difficult to distinguish 
from the strong H$_2$O and CH$_4$ bands and so until even later-type dwarfs 
are available to define the Y  class we adopt a type of T9 here
(as also suggested by Burningham et al.).

The three dwarfs classified as T9 have narrower  $J$ and $H$ (1.2 and 1.6~$\mu m$)
flux peaks than  2MASS J0415$-$09,  suggesting that they are  cooler than 750~K.
The near-infrared indices, including the ``W$_J$'' and ``NH$_3$-$H$'' indices,
confirm an earlier type of T8 for 2MASS J0939$-$24.
However, as shown by Burgasser et al. (2008) and as we demonstrate below, the 
mid-infrared data for 2MASS J0939$-$24 constrain its temperature to be similar to the 
T9 dwarfs. The near-infrared spectrum of 2MASS J0939$-$24 is very low resolution with 
$R \approx 150$ or $\Delta\lambda\approx 0.01 \mu$m, so that detection of the
variation in the width of the $J$ and $H$ bands is somewhat compromised; however 
the object is further complicated by possible binarity (Burgasser et al. and \S 5.4)
and we show later that multiplicity can addresses the discrepancy between 
near-infrared spectral type and temperature. 
 
Table 1 lists astrometric and photometric data for the three T9 dwarfs and 2MASS J0939$-$24, taken from the sources 
referenced in the Table. The $z$ photometry for the UKIDSS dwarfs has been converted to the SDSS AB system using 
transformations given by Warren et al. (2007). The $YJHK$ photometry is on the Mauna Kea Observatories Vega-based 
system (Tokunaga et al. 2002; the $JHK$ for 2MASS J0939$-$24 is synthesized from its spectrum which was flux 
calibrated using 2MASS photometry). The IRAC photometry is reported in the standard IRAC system, which is calibrated 
against Vega.  The IRAC values provide a flux density at a nominal wavelength assuming a flux distribution where $\nu 
f_{\nu}$ is constant, which is not the case for T dwarfs.  Hence the IRAC magnitudes do not give the flux at the 
nominal wavelengths listed in Table 1, but do give a scaling factor for the spectrum over the IRAC filter bandpasses 
(e.g. Cushing et al. 2006).

\section{{\em Spitzer} IRS Observations}

We used the Short-Low (SL) module of the IRS to obtain first-order 7.6 -- 15.2~$\mu m$ spectra of ULAS J0034$-$00 and 
ULAS J1335$+$11 in the standard staring mode, which produces low-resolution spectra with $\lambda/\Delta\lambda 
\approx$ 60 -- 120. Wavelengths longer than $\sim 14.5 \mu$m are compromised by an overlapping second order. 
The blue peakup band was used to center the targets in the 3.7'' slit by offsetting from a nearby 
bright star, and cycles were repeated after nodding the target along the slit by roughly one- and two-thirds of its 
length.

ULAS J0034$-$00 was observed as part of program 40419 using instrument-team guaranteed time in Cycle 4. The dwarf was observed on 2008 January 15 and January 16, each time using a 240 second ramp duration and 32 cycles. The total time on source was 8.5 hours, and the total AOR
duration was 10.8 hours.

The brighter dwarf ULAS J1335$+$11 was observed as part of program 40449 using general observer time in Cycle 4. The dwarf was observed on 2008 August 9, using a 60 second ramp duration and 64 cycles. The total time on source was 2.1 hours, and the total AOR duration was 2.8 hours.

Low-level processing of the spectra was done automatically by versions 17.2.0 and 18.1.0 
of the IRS pipeline for ULAS J0034$-$00 and ULAS J1335$+$11, respectively. 
The IRS pipeline removed all of the instrumental artifacts to produce Basic Calibrated Data
(BCD) for each exposure.
The data were further reduced as described in Cushing et al. (2006).  
Briefly, the spectra were
extracted using a modified version of the Spextool package with a fixed
aperture of 6''.  Observations of $\alpha$ Lac, obtained as part of the
IRS calibration observations, were used to remove the instrument response
and flux calibrate the spectra.
The flux calibration was checked using the IRAC band 4 (6.4 -- 9.3 $\mu$m) photometry,
extrapolating the observed spectrum bluewards from 7.6 to 6.4 $\mu$m using the observed
spectrum of the T8 dwarf 2MASS J0415$-$09 as a template. Very little flux is emitted
at these wavelengths due to strong absorption by H$_2$O at $\sim$5 -- 7 $\mu$m and CH$_4$
at $\sim$7 -- 9 $\mu$m. This photometric check showed that the IRS calibration is
accurate to $\leq$10\%.

Figure 2 shows the IRS spectra of ULAS J0034$-$00 and ULAS J1335$+$11.

\section{Atmospheric and Evolutionary Models}

Atmospheric and evolutionary models were generated by members of our team (Marley et al. 2002, Saumon \& Marley 
2008). The atmospheric models include all the significant sources of gas opacity (Freedman et al.\ 2008). However 
there are known deficiencies in the molecular opacity line lists at $0.9 < \lambda \ \mu$m $< 1.4$, where the line 
list for CH$_4$ is incomplete and the line list for NH$_3$ does not exist. At the temperatures considered here these 
opacities are significant and we return to this issue later in \S 5 and \S 6. The models assume solar system 
abundances and calculate the equilibrium abundances of the important C, N, and O bearing gases as detailed by 
Freedman et al.\ (2008).  These equilibrium values are then used to determine the abundances of the various gas 
species in each atmospheric layer during construction of the temperature-pressure profile. 
We iteratively compute models in
radiative-convective equilibrium, solving for a temperature-pressure profile while employing the source-function
technique of Toon et al. (1989) to solve the equation of transfer.
During this process, condensates of certain refractory elements are 
included in atmospheric layers where physical conditions favor such condensation.  At the temperatures considered 
here the condensate clouds are located deep below the photosphere and the opacity of the clouds is ignored although 
grain condensation is still accounted for in the chemistry.

The models also include vertical transport in the atmosphere, which significantly
affects the chemical abundances of C-, N- and O-bearing species.  
The very stable molecules CO and N$_2$ are dredged up
from deep layers to the photosphere, causing
enhanced abundances of these species and decreased abundances of CH$_4$, H$_2$O, and
NH$_3$ (e.g. Fegley \& Lodders 1996,  Hubeny \& Burrows 2007, Saumon et al. 2007).  
Vertical transport is parameterized in our models by
an eddy diffusion coefficient $K_{zz}$ (cm$^{2}$ s$^{-1}$) that is related to the
mixing time scale. Generally, larger values of $K_{zz}$ imply greater enhancement
of CO and N$_2$ over CH$_4$ and NH$_3$, respectively.  Values of $\log K_{zz} = 2$ -- 6,
corresponding to mixing time scales of $\sim 10$~yr to $\sim 1$~h respectively, reproduce the
observations of T dwarfs (e.g. Leggett et al. 2007a, Saumon et al. 2007).

The evolutionary sequences of Saumon \& Marley (2008), computed with cloudless models as the surface-boundary 
condition, provide radius, mass, and age for a variety of atmospheric parameters. For this investigation synthetic 
spectra were generated with $T_{\rm eff}$ of 500, 550, 600, 650 and 700~K; gravity $g$ of 100, 300, 1000 and 2000 
ms$^{-2}$ (corresponding to $\log g$ in c.g.s. units of 4.00, 4.48, 5.00 and 5.30); 
metallicities of solar and twice solar ([m/H]$= +0.3$); and vertical mixing diffusion coefficient $K_{zz}$ 
of zero (corresponding to equilibrium) and $10^4$ cm$^2$s$^{-1}$. A limited set was also calculated with 
$K_{zz}=10^6$ cm$^2$s$^{-1}$ and [m/H]$= -0.3$. The evolutionary models show that a brown dwarf with $T_{\rm eff} 
\leq 700$ K must have $g \leq 2000$ ms$^{-2}$ if its age is $\leq 10$ Gyr (Saumon \& Marley 2008).

Figure 3 plots a selection of 500~K and 600~K synthetic spectra to demonstrate the effects
of varying the atmospheric parameters.  The top panel shows that metallicity and gravity variations
can affect the spectral energy distributions similarly --- an increase in metallicity or a
decrease in gravity leads to significantly increased flux in the $K$-band at 2.2 $\mu$m, and decreased flux at 4.5 $\mu$m (impacting IRAC band 2 photometry for example).  The opposite changes occur if metallicity is decreased or gravity increased. The lower panel shows that varying $T_{\rm eff}$ impacts the overall slope of the spectral energy distribution --- decreasing $T_{\rm eff}$ increases the ratio of the mid- to near-infrared fluxes (although surface flux at all wavelengths decreases with decreasing temperature). The lower panel also shows that the mixing parameter
$K_{zz}$ strongly affects the 4.5 $\mu$m and 9 -- 15 $\mu$m fluxes:  increasing $K_{zz}$  reduces the 4.5 $\mu$m flux as CO is dredged up, and increases the 9 -- 15 $\mu$m flux as N$_2$ is favored over NH$_3$. For the objects considered here the CH$_4$ and H$_2$O bands are 
not significantly affected by CO mixing, as the amount of carbon and oxygen tied up in CO is a small 
fraction of that in these other molecules at these temperatures.

\section{Analysis}

\subsection{The  0.9 -- 2.5 $\mu$m and 7.6 -- 14.5 $\mu$m Spectra of ULAS J0034$-$00 and ULAS J1335$+$11}

The synthetic spectra were compared to the observed 0.9 -- 2.5 $\mu$m and 7.6 -- 14.5 $\mu$m
spectra of ULAS J0034$-$00 and ULAS J1335$+$11 using the fitting procedure described by Cushing et 
al.\ (2008), except that here we do not mask out the 1.58
to 1.75 $\mu$m spectral region. No regions were ignored in the fitting, despite the known problems 
at $0.9 < \lambda \ \mu$m $< 1.4$ (\S 4). The procedure uses weighted least-squares fitting to
identify the synthetic spectrum that best matches the observed spectrum.
The scaling of the  synthetic spectra to the observations is determined as part of the fitting procedure,
by minimising the weighted deviation between the two datasets.


Fits using the entire wavelength range as well as only the near- or mid-infrared were carried out. 
For both of these T9 dwarfs the fits to either the near- or mid-infrared regions alone imply a value of $T_{\rm eff}$ 
of $600 \leq T_{\rm eff} \leq 700$. However the fit to the mid-infrared spectrum 
does not constrain the temperature strongly, and the fit to the near-infrared is very poor in the $H$-band.  The fits 
to the full spectrum gives a lower temperature of $500 \leq T_{\rm eff} \leq 600$ for ULAS J0034$-$00 and $500 \leq 
T_{\rm eff} \leq 550$ for ULAS J1335$+$11; in these full-range fits the match to the $H$-band is improved at the 
expense of a poorer fit to the $K$-band. Increasing $T_{\rm eff}$ makes the overall spectrum bluer (e.g. Figure 3), 
and temperatures as high as suggested by the near-infrared fits cannot reproduce the observed full spectral energy 
distributions.

The best fitting model to the spectrum of ULAS J0034$-$00 has 
effective temperature (K) / gravity (ms$^{-2}$) / metallicity ([m/H] dex)  of
550 / 100 / 0.0; this fit is shown in the top panel of Figure 4. However temperatures in the range 
550 -- 600~K, gravities of  100 -- 300 ms$^{-2}$ and metallicities of 0.0 -- 0.3 dex do not
give significantly worse fits, as shown in the top two panels of Figures 4 and 5.
The best fitting model to the spectrum of ULAS J1335$+$11 has 
effective temperature (K) / gravity (ms$^{-2}$) / metallicity ([m/H] dex)  of
500 / 100 / 0.0; this fit is shown in the top panel of Figure 6.
Temperatures in the range 
500 -- 550~K, gravities of  100 -- 300 ms$^{-2}$ and metallicities of 0.0 -- 0.3 dex
give similar quality fits as shown in Figure 6.

The full spectral range supports a low gravity for both dwarfs of $100 \leq g$ \ ms$^{-2}$ $\leq 300$. Although an 
increase in gravity can be partly compensated for by increasing metallicity, a gravity as high as $g = 1000$ 
ms$^{-2}$ gives a poorer least-squares fit to both dwarfs, and this can be seen for ULAS 
J0034$-$00 in the bottom panel of Figure 4.

For both dwarfs, models which include gas transport, i.e. non-equilibrium chemistry with $K_{zz} > 0$, are strongly 
preferred by the fits to the mid-infrared only and the fits to the entire spectral range. The equilibrium chemistry 
models do not reproduce the shape of the 9 -- 15 $\mu m$ spectrum, where NH$_3$ absorption is strong, but not as 
strong as it would be in the $K_{zz} = 0$ case; this is shown in the bottom panel of Figure 5. The flux at 4.5 $\mu$m 
is also dependent on the value of $K_{zz}$, which we discuss further when considering the IRAC photometry below.
Vertical mixing in ULAS J1335$+$11 appears stronger than in ULAS J0034$-$00 based on the 4.5 $\mu$m IRAC photometry; 
we determine $K_{zz} \approx 10^6$ cm$^2$s$^{-1}$ cf. $10^4$ cm$^2$s$^{-1}$ for ULAS J0034$-$00 (\S 5.2).

\subsection{The 3 - 9 $\mu$m Photometry of ULAS J0034$-$00 and ULAS J1335$+$11}

The IRAC photometry for ULAS J0034$-$00 and ULAS J1335$+$11 (Warren et al. 2007, 
Burningham et al. 2008) fills the gap between the near- and mid-infrared spectra
and provides another check of the model fit. The observed and calculated IRAC fluxes 
are indicated in Figures 4, 5 and 6. The observed fluxes are determined by transforming the observed magnitudes to fluxes using the Vega fluxes of 280.9, 179.7, 115.0 and 64.1 Jy at the
four IRAC bands [3.6], [4.5], [5.8] and [8.0], respectively, as provided in the  IRAC Data Handbook.  The calculated flux is determined in the same way from synthetic absolute magnitudes, scaling by the distance to the dwarf, $D$, which is provided by the spectral fit.
The scaling factor required to match the synthetic spectra to the data is a function
of the solid angle $(R/D)$ where $R$ is the radius of the dwarf. As $R$ is known from the evolutionary model corresponding to the particular set of atmospheric parameters $T_{\rm eff}$, $g$ and [m/H], the distance $D$ can be derived. Both dwarfs are assumed to be single in this analysis; both have been imaged at high resolution and not resolved (N. Lodieu and M. Liu,
private communications).

The IRAC [4.5] flux is sensitive to the vertical transport coefficient $K_{zz}$ as shown in Figures 3 and 5. 
Increasing $\log K_{zz}$ from 4 to 6 reduces the [4.5] flux by 10 -- 25\% at these temperatures and gravities, as 
the non-equilibrium abundance of CO increases.
The other IRAC bands are insensitive to $K_{zz}$ at these low temperatures, as described in \S 4 and 
shown in the middle panel of Figure 5. Hence $\sim 20$\% effects at [4.5] can be neglected when examining the model 
fits as these can be addressed by varying $K_{zz}$.

All the models preferred by the spectral fits to ULAS J0034$-$00 give IRAC [3.6], [5.8] and [8.0] fluxes which are 
too low by 30 -- 50 \%. The agreement at [4.5] is good as long as $K_{zz} > 0$ (which is required by the spectral 
fits also). The fits to the [3.6], [5.8] and [8.0] fluxes are better for ULAS J1335$+$11, while the [4.5] model flux 
is high by 20 -- 30 \% for this dwarf, even with $K_{zz}= 10^6$ cm$^2$s$^{-1}$. The fact that subsets of the modelled 
and observed IRAC photometry agree well for each dwarf suggests that a more detailed model grid and a model fitting 
routine which incorporates both the spectroscopy and photometry could provide a more consistent set of observed and 
synthetic data. We postpone such work until the molecular line lists are improved (see \S 6).

Despite the remaining $\sim$30\% discrepancies at [3.6], [5.8] and [8.0] for ULAS J0034$-$00 (regions where there is 
very little flux), the near- and mid-infrared spectra, and the IRAC photometry, together with our models, can 
satisfactorily constrain $T_{\rm eff}$ (K)/$g$ (ms$^{-2}$)/[m/H] (dex) to 550 -- 600/100 -- 300/0.0 
-- 0.3 and 500 -- 550/100 -- 300/0.0 -- 0.3 for ULAS J0034$-$00 and ULAS J1335$+$11, respectively. Vertical transport 
of gas must be significant, with a diffusion coefficient of $K_{zz} = 10^4$ -- $10^6$ cm$^{2}$ s$^{-1}$, similar to 
values found for earlier-type T dwarfs.

These parameters and their corresponding radii, masses, ages, distances and tangential velocities are given in Table 
2. The low temperatures and gravities imply that the dwarfs have masses of only 5 -- 20 M$_{Jupiter}$ and young ages 
of 0.1 -- 2 Gyr. We discuss the uncertainties in these parameters in \S 6. We note that the implied distance of 13 -- 
16 pc for ULAS J0034$-$00 is in excellent agreement with a preliminary trigonometric parallax of 14 $\pm$ 3 pc (R. 
Smart, private communication).

\subsection{Trends in Color}

Fitting the synthetic spectra to the observations showed that the ratio of the
mid- to near-infrared flux is very sensitive to temperature.  Combining ground-based
near-infrared photometry with IRAC photometry therefore should be a useful indicator
of  $T_{\rm eff}$, in particular the [4.5] band as there is very little flux at
[3.6], [5.8] and [8.0].  Warren et al. (2007) investigate various color combinations 
and find that $H -$[4.5] is a good indicator of $T_{\rm eff}$ for T dwarfs. 
While the $K$-band is sensitive to metallicity and gravity, the $H$-band is insensitive
(e.g. Figure 3).
Stephens et al. (2009) show that T dwarf colors involving $J$ and IRAC magnitudes
show a lot of scatter with $T_{\rm eff}$, at least some of which is due to condensates
clearing out of the photospheric region probed by 1.2 $\mu$m radiation in the earlier
T dwarfs.
   
Figure 7 shows  $H -$[4.5] as a function of $H - K$. 
Model sequences are shown for a range of gravity and metallicity, and for non-equilibrium chemistry
models with $K_{zz} = 10^4$ cm$^{2}$ s$^{-1}$. Increasing $K_{zz}$ by a factor of 100
leads to no significant change in $H - K$, but a decrease in   $H -$[4.5] of 0.1 -- 0.3
magnitudes.  From this and other studies it appears that the range of  $K_{zz}$ in T dwarfs
is $10^2$ -- $10^6$ cm$^{2}$ s$^{-1}$ and so the uncertainty in the modelled  $H -$[4.5]
due to unknown  $K_{zz}$ should be $\approx$ 0.2 magnitudes.
Plausible changes in gravity and metallicity for a local disk sample also affect
 $H -$[4.5] by $\sim$0.2 magnitudes and affect 
$H - K$ by $\sim$0.3 -- 0.6 magnitudes at the temperatures considered here 
(note the effect on $H - K$ increases with decreasing temperature). 
As described before, the trends with metallicity and gravity are degenerate.
Although these  $\sim$0.2 magnitude uncertainties are not ideal in a temperature indicator,
Figure 7 shows that for  $T_{\rm eff} < 700$ K  $H -$[4.5] increases by 0.6 -- 0.8
magnitudes per 100~K and so temperatures can still be accurately determined.
The {\it Spitzer} warm mission retains IRAC [4.5] capability and hence will continue
to provide important data for studies of brown dwarfs.

Colors for ULAS J0034$-$00, ULAS J1335$+$11 and 2MASS J0939$-$24 are shown in Figure 7, as well as those of other 
dwarfs taken from the literature (Patten et al. 2006, Leggett et al. 2007a, Luhman et al. 2007). The T7.5 dwarfs Gl 
570D and HD 3651B are benchmark dwarfs with well constrained ages and metallicities from their primary stars, and 
well constrained $T_{\rm eff}$ based on their measured luminosity and the known distance to the primary star 
(Burgasser et al. 2000, Burgasser 2007, Liu et al. 2007, Mugrauer et al. 2006, Saumon et al. 2006). Gl 570D has 
$T_{\rm eff}=$ 750 -- 825 K, $g =$ 630 -- 2000 ms$^{-2}$ and [m/H] $=$ 0.01 -- 0.10; HD 3651B has $T_{\rm eff}=$ 780 
-- 840 K, $g =$ 1260 -- 3160 ms$^{-2}$ and [m/H] $=$ 0.09 -- 0.16. 2MASS J0415$-$09 has a well measured distance and 
spectral energy distribution which indicates $T_{\rm eff} =$ 725 -- 775~K, $g =$ 1000 -- 2500 ms$^{-2}$ and [m/H] $=$ 
0.0 -- 0.3 dex (Saumon et al. 2007). If the sequences shown in Figure 7 are shifted redwards in $H - K$ by $\sim$0.2 
magnitudes and redwards in $H -$[4.5] by $\sim$0.1 magnitudes there is good agreement with the observations of these 
slightly metal-rich dwarfs, and with the overall population in the diagram. The models appear to be too faint at $K$, 
and this can also be seen in Figures 4 through 6. Most likely this is due to some inadequacy in the CH$_4$ line list 
or the H$_2$ collision-induced absorption, as discussed in Saumon \& Marley (2008; their \S 4.1).

The color trends shown in  Figure 7 support the temperatures, gravities and metallicities found above for 
ULAS J0034$-$00 and ULAS J1335$+$11: that they are relatively low-gravity dwarf with solar or enhanced
metallicities, and that ULAS J1335$+$11 is cooler than ULAS J0034$-$00.  
The colors also support the conclusion of Burgasser et al.
(2008) in their analysis of the T8 dwarf 2MASS J0939$-$24: that it has a similar 
temperature to these T9 dwarfs, but a higher gravity and/or lower metallicity.
We explore fits to this object's spectral distribution in the following section.

\subsection{
Fitting the Spectral Energy Distribution of 2MASS J0939$-$24}

Burgasser et al. (2008) find the following parameters for 2MASS J0939$-$24 using 
near- through mid-infrared observational data:
$T_{\rm eff}= 600 \pm 50$ K, $g =$ 600 ms$^{-2}$ and 
[m/H] $= -0.4$ (the uncertainties in $g$ and [m/H] are quite large). 
The same model set and a similar least-squares fitting routine were
used as we use here for ULAS J0034$-$00 and ULAS J1335$+$11. In this case however the
distance to the dwarf is known and so the scaling factor required to match the 
models to the observations gives the radius of the dwarf, or alternatively if the radius
is known from evolutionary arguments then the scaling factor can be derived.
Adopting the first approach Burgasser et al. show that the radius is larger
than expected from the models, and they suggest the dwarf is a binary.

Figure 8 shows this dwarf's observed spectroscopy, IRAC photometry and a selection of modelled data. We have selected 
models with $T_{\rm eff}= 600$ K and values of gravity and metallicity within the range determined by Burgasser et 
al.'s least-squares fitting. Scaling of the modelled spectra and photometry is achieved using the known distance to 
the dwarf and the radius from evolutionary models. We assume that either it is single or an equal luminosity 
unresolved pair, as best matches the data. The lower gravity fit, which would be marginally consistent with this 
object being single, is a significantly poorer fit to the spectra and photometry than the higher gravity fits,
which require it to be a binary system.  Also 
a warmer model with $T_{\rm eff}=$ 650 K can reproduce the near-infrared data well if the dwarf is single, but is 
under-luminous in the mid-infrared by $\sim$30\%.  A reasonable fit can be found if indeed, as suggested by Burgasser 
et al., this is a pair of similar luminosity 600~K dwarfs with $g \approx 1000$ ms$^{-2}$ and [m/H] $= -0.3$ -- 0.0 
dex.

Given that a temperature as low as 600 K is unexpected for an object with the same
near-infrared spectral type as the 750 K dwarf 2MASS J0415$-$09, and that the near-infrared
data led Leggett et al. (2007b) to assign $T_{\rm eff} = 750$ K to 2MASS J0939$-$24,
we explored the possibility that this object consists of a more dissimilar pair of dwarfs.
We find that a satisfactory fit is produced if 2MASS J0939$-$24 is composed of a
700 K and a 500 K dwarf. The gravities must be high so that the radii are small, otherwise 
the total flux  of the system is too large.  The relative gravities are also constrained
by the evolutionary models, as the dwarfs presumably are coeval. Fits with 
$g =$ 1000  ms$^{-2}$ for the 500 K dwarf constrains the gravity of the 700 K dwarf to
1000 -- 2000 ms$^{-2}$.  The system still appears to be somewhat metal-poor with [m/H] 
$= -0.3$ -- 0.0 dex. Figure 9 shows composite fits for this range of parameters.
To improve the agreement with the previously determined temperature of this object,
and to make the temperature:spectral type correlation stronger, this
dissimilar-pair solution is preferred to the equal-luminosity pair solution.
The near-infrared spectrum is then dominated by the warmer dwarf (see Figure 9)
and becomes consistent with the T8 spectral type. The modelled dissimilar pair has slightly
wider $J$ and $H$ flux peaks than the equal luminosity pair, consistent with the 
wider peaks observed for 2MASS J0939$-$24 and 2MASS J0415$-$09 compared to
the three T9 dwarfs (Figure 1). 

However the detailed spectral shape is sensitive to gravity and metallicity as well as 
temperature, and the temperature:spectral type relationship will have some scatter. Hence
unless the object is resolved it cannot be determined whether the equal or non-equal pair
solution is correct.  The physical properties of both solutions
are listed in Table 2. The 500 K $+$ 700 K pair are constrained to be high gravity and 
therefore quite old at 4 -- 10~Gyr; perhaps appropriate for the slightly metal-poor
nature of the system.

\subsection{The 0.9 -- 2.5 $\mu$m Spectrum of CFBD J0059$-$01}

At this time there are no mid-infrared data published for the T9 dwarf CFBD J0059$-$01.
Figure 1 shows that the near-infrared spectrum of CFBD J0059$-$01 is very similar to 
ULAS J0034$-$00 and ULAS J1335$+$11, except that the $K$-band flux is suppressed.
Burningham et al. (2008) show that the ``W$_J$'', ``NH$_3$-$H$'' and ``H$_2$O-$H$'' 
indices place CFBD J0059$-$01 consistently between ULAS J0034$-$00 and ULAS J1335$+$11,
and Delorme et al. (2008a) conclude that it is $\sim$50~K cooler than  ULAS J0034$-$00
based on a comparison of near-infrared features. The indices and $K$ flux  suggest that  
CFBD J0059$-$01 has a temperature between  that of ULAS J0034$-$00 and ULAS J1335$+$11
and either a higher gravity or lower metallicity. The $H - K$ value is intermediate between the
ULAS T9 dwarfs and 2MASS J0939$-$24 (see Figure 6), suggesting that either it is not as metal-poor or 
as high gravity as the latter dwarf (or pair of dwarfs). Delorme et al. find that CFBD J0059$-$01 is 
unresolved at high resolution and so here we assume it is single. 

Fitting the near-infrared spectrum alone, we find  that a temperature as low as 500~K
does not give a satisfactory fit; $T_{\rm eff}=$ 550 -- 600 K produces similar fits
in the near-infrared as the other dwarfs, when  $g =$ 300 -- 2000  ms$^{-2}$ and 
[m/H] $=$ 0 -- 0.3 dex. The high metallicities are required if $T_{\rm eff}=$ 550 K 
and $g =$ 1000 -- 2000  ms$^{-2}$, to fit the $K$-band flux. 
This temperature range of 550 -- 600 K is in agreement with the trends in the  near-infrared
indices, and Delorme et al.'s conclusion that CFBD J0059$-$01 is cooler than
ULAS J0034$-$00, if CFBD J0059$-$01 is at the low-end of the range while 
ULAS J0034$-$00 is at the high end.

The range of implied distances and other physical
properties for  CFBD J0059$-$01 are given in Table 2. This dwarf may be significantly older and more
massive than ULAS J0034$-$00 and ULAS J1335$+$11; its tangential velocity is also
significantly higher ($\sim$50 kms$^{-1}$ cf. 25 kms$^{-1}$). Mid-infrared data will greatly help
constrain its parameters. Such data are needed to confirm the temperature implications of
the near-infrared indices, and will also constrain metallicity and gravity by fixing
the scaling of the models and restricting the slope of the near-infrared flux distribution.


\section{Discussion and Conclusions}

We have used new and published near- through mid-infrared spectroscopy and photometry to  
constrain the atmospheric parameters of the very low-temperature dwarfs 
ULAS J0034$-$00, ULAS J1335$+$11 and 2MASS J0939$-$24. 
We find strong evidence that photospheric mixing continues to be an important process, even at
these cool temperatures.  Any 5 $\mu$m survey for Y dwarfs (e.g. WISE or Spitzer warm missions) 
must take this mixing into account, as the flux will be less than predicted by equilibrium chemistry. 
Considering the complexity of the atmospheres, the models reproduce the data well.
However there are known deficiencies in the molecular opacity line lists, and discrepancies
can be seen in Figures 4 -- 6 and 8.  The known problems are large at $0.9 < \lambda \ \mu$m
$< 1.4$, where the line list for CH$_4$ is incomplete, the line list for NH$_3$ does not exist,
and where treatment of the red wing of the very broad 0.77 $\mu$m K~I line is difficult. These
problems most likely account for the excess model flux seen in our fits at $1.0 < \lambda \ \mu$m
$< 1.3$.

Given these deficiencies, the systematic errors in our derived parameters are difficult to determine. However, the 
good agreement with evolutionary models for the $\sim$800 K brown dwarf companions to stars of known age and 
metallicity 
means that it is unlikely that the errors are larger than 100 K in $T_{\rm eff}$, or a factor of two in gravity and 
metallicity (see the trends shown in Figures 3 and 7). Nevertheless it is of paramount importance that the line lists 
for CH$_4$ and NH$_3$ are improved, now that objects as cool as 500 K are being discovered.

It is also very helpful at these low temperatures to have mid-infrared observational data. The absorption features in 
the near-infrared are effectively saturated, and the changes with temperature have become subtle. The mid-infrared 
flux level is very sensitive to temperature and can remove this degeneracy; this sensitivity increases with 
decreasing temperature as shown in Figure 7. Mid-infrared data offers the possibility of identifying low-temperature 
systems such as 2MASS J0939$-$24, which may be a 500 K $+$ 700 K brown dwarf pair. We find that two of the three 
currently known T9 dwarfs have $ 550 \leq T_{\rm eff}$ K $\leq 600$ and the third appears to be even cooler with $ 
500 \leq T_{\rm eff}$ K $\leq 550$. Vertical transport of gas remains significant even at these very low 
temperatures, so that enhanced CO opacity reduces the flux at 4.5~$\mu$m and decreased NH$_3$ opacity increases the 
flux at 9 -- 15 $\mu$m. 

The evolutionary models show that the parameters we derive for ULAS J0034$-$00 and 
ULAS J1335$+$11 imply that they are both very low mass and relatively young ---
their masses are between 5 and 20 M$_{\rm Jupiter}$ corresponding to an age range of
0.1 to 2.0 Gyr. The ranges for CFBD J0059$-$01 are larger --- 10 -- 50 M$_{\rm Jupiter}$
and 0.5 -- 10 Gyr.  If 2MASS J0939$-$24 consists of a 500 K $+$ 700 K binary
then it is relatively old and massive: a 4 -- 10 Gyr system composed of a 
25 M$_{\rm Jupiter}$ and a 40 M$_{\rm Jupiter}$ brown dwarf pair.


\ack

This work is based  on observations made with the {\it Spitzer Space Telescope}, which is operated by the Jet Propulsion Laboratory, California Institute of Technology under a contract with NASA. Support for this work was provided by NASA through an award issued by JPL/Caltech.
SKL's research is supported by the Gemini Observatory, which is operated by the Association of Universities for Research in Astronomy, Inc., on behalf of the international Gemini partnership of Argentina, Australia, Brazil, Canada, Chile, the United Kingdom, and the United States of America.

\clearpage

\References


\item[] Burgasser A et al. \ 2000 {\it ApJ} {\bf 531} L57

\item[] Burgasser A et al. \ 2002 {\it ApJ} {\bf 564} 421

\item[] Burgasser A, Geballe T R, Leggett S K, Kirkpatrick J D and Golimowski D A \ 2006 
{\it ApJ} {\bf 637} 1067

\item[] Burgasser A\ 2007 {\it ApJ} {\bf 658} 617

\item[] Burgasser A, Tinney C G, Cushing M C, Saumon D, Marley M S, Bennett C S and 
Kirkpatrick J D \ 2008, {\it ApJ} {\bf 689} L53

\item[] Burningham B et al. \ 2008, {\it MNRAS} {\bf 391} 320

\item[] Burrows A, Sudarsky D, and Lunine J I \ 2003  {\it ApJ} {\bf 596} 587


\item[] Cushing M C et al. \ 2006 {\it ApJ} {\bf 648} 614

\item[] Cushing M C et al. \ 2008 {\it ApJ} {\bf 678} 1372

\item[] Delorme P et al. \ 2008 {\it A\&A} {\bf 482} 961

\item[] Delorme P et al. \ 2008 {\it A\&A} {\bf 484} 469

\item[] Epchtein N \ 2008 {\it Proc. 179th Symposium of the International Astronomical Union}
(Kluwer Academic Publishers) p 106

\item[] Fazio, G. et al. \ 2004 {\it ApJS} {\bf 154} 10

\item[] Fegley B and Lodders K \ 1996 {\it ApJL} {\bf 472} L37

\item[] Freedman R S, Marley M S and Lodders K \ 2008  {\it ApJS} {\bf 174} 504

\item[] Golimowski D A et al.\ 2004 {\it AJ} {\bf 127} 3516

\item[] Houck J R et al. \ 2004 {\it ApJS} {\bf 154} 18

\item[] Hubeny I and Burrows A \ 2007 {\it ApJ}  {\bf 669} 1248


\item[] Kirkpatrick D J, Beichman C A and Skrutskie M F \ 1997 {\it ApJ} {\bf 476} 311

\item[] Kirkpatrick D J et al. \ 1999 {\it ApJ} {\bf 519} 802

\item[] Knapp G R et al. \ 2004 {\it AJ} {\bf 127} 3553

\item[] Lawrence A et al. \ 2007 {\it MNRAS} {\bf 379} 1599

\item[] Leggett S K et al. \ 2000 {\it ApJ} {\bf 536} L35

\item[] Leggett S K, Saumon D, Marley M S, Geballe T R, Golimowski D A, Stephens D and 
Fan X \ 2007a {\it ApJ} {\bf 655} 1079

\item[] Leggett S K, Marley M S, Freedman R, Saumon D,  Liu M C, Geballe T R, Golimowski D A, and Stephens D  \ 2007b {\it ApJ} {\bf 667} 537

\item[] Liu M C, Leggett S K and Chiu K  \ 2007 {\it ApJ} {\bf 660} 1507


\item[] Luhman K L et al. \ 2007 {\it ApJ} {\bf 654}  570

\item[] Marley M S, Seager S, Saumon D, Lodders K, Ackerman A S, Freedman R S,
  and Fan X \ 2002  {\it ApJ} {\bf 568} 335

\item[] Martin E L, Delfosse X, Basri G, Goldman B, Forveille T and 
Zapatero Osorio M \ 1999  {\it AJ} {\bf 118} 2466

\item[] McLean I et al. \ 2003 {\it ApJ} {\bf 596} 561

\item[] Mugrauer M, Seifahrt A, Neuhauser R and Mazeh T \ 2006 
{\it MNRAS} {\bf 373} L31

\item[] Patten B M et al. \ 2006 {\it ApJ} {\bf 651} 502

\item[] Saumon D, Marley M S, Cushing M C, Leggett S K, Roellig T L, Lodders K
and Freedman R S \ 2006 {\it ApJ} {\bf 647} 552

\item[] Saumon D et al. \ 2007 {\it ApJ} {\bf 662} 1245

\item[] Saumon D and Marley M S \ 2008 {\it ApJ} 


\item[] Skrutskie M F et al.\ 2006 {\it AJ} {\bf 131} 1163


\item[] Stephens D S, Leggett S K, Cushing M C, Saumon D, Marley M S, Geballe T,
Golimowski D A and Fan X \ 2009, {\it ApJ}

\item[] Strauss M A et al.\ 1999 {\it ApJ} {\bf 522} 61

\item[]  Tinney C G, Burgasser A J, Kirkpatrick J D and McElwain M W \ 2005
 {\it AJ} {\bf 130} 2326

\item[]  Tokunaga A T, Simons D A and Vacca W D \ 2002 {\it PASP} {\bf 114} 180

\item[]  Toon O B, McKay C P, Ackerman T P and Santhanam K \ 1989 {\it JGR}
{\bf 94} 16287

\item[] Vrba F J et al.\ 2004 {\it AJ} {\bf 127} 2948


\item[] Warren S J et al. \ 2007 {\it MNRAS} {\bf 381} 1400

\item[] Werner M W et al.\ 2004 {\it ApJS} {\bf 154} 1

\item[] York D G et al.\ 2000 {\it AJ} {\bf 120} 1579

\endrefs

\clearpage



\begin{table}
\small
\caption{\label{tab1}Astrometry and  Photometry of the Sample.}
\begin{tabular}{@{}lrrrr}
\br
Property & ULAS J0034$-$00$^{\rm a}$ & CFBD J0059$-$01 & 2MASS J0939$-$24$^{\rm b}$ & ULAS J1335$+$11\\
\mr
Right Ascension & 00:34:02.77 & 00:59:10.90 & 09:39:35.48 & 13:35:53.45  \\
Declination & $-$00:52:06.7 & $-$01:14:01.1 & $-$24:48:27.9 & $+$11:30:05.2 \\
Epoch yyyymmdd & 20051004 & 20070805 & 20000210 & 20070421 \\
$\mu_{\alpha}$ "/yr & $-0.12\pm$0.05 & 0.94$\pm$0.06 & 0.573$\pm$0.002 & 0.16$\pm$0.05 \\
$\mu_{\delta}$ "/yr & $-0.35\pm$0.05 & 0.18$\pm$0.06 & $-$1.045$\pm$0.003 & $-0.29\pm$0.05 \\
Distance pc & 14 $\pm$ 3   &    &  5.34$\pm$0.13 & \\
$z_{\rm AB}$ & 22.11$\pm$0.05 & 21.93$\pm$0.05 & & 22.04$\pm$0.10  \\
$Y$ & 18.90$\pm$0.10 & 18.82$\pm$0.02 & & 18.81$\pm$0.04  \\
$J$ & 18.15$\pm$0.03 & 18.06$\pm$0.03 & 15.61$\pm$0.09  & 17.90$\pm$0.01 \\
$H$ & 18.49$\pm$0.04 & 18.27$\pm$0.05 & 15.96$\pm$0.09 & 18.25$\pm$0.01 \\
$K$ & 18.48$\pm$0.05 & 18.63$\pm$0.05 & 16.83$\pm$0.09 & 18.28$\pm$0.03 \\
3.55 $\mu$m & 16.28$\pm$0.03 & & 13.76$\pm$0.04 & 15.96$\pm$0.03 \\
4.49 $\mu$m & 14.49$\pm$0.03 & & 11.66$\pm$0.04 & 13.91$\pm$0.03 \\
5.73 $\mu$m &  14.82$\pm$0.05 & & 12.96$\pm$0.04 & 14.34$\pm$0.05 \\
7.87 $\mu$m & 13.91$\pm$0.06  &  & 11.89$\pm$0.04 & 13.37$\pm$0.07 \\
References & 1 & 2 & 3, 4 & 5 \\
\br
\end{tabular}
\noindent
$^{\rm a}$ The preliminary parallax is a private communication from R. Smart.\\
$^{\rm b}$ The MKO-system $JHK$ photometry is synthesized from the spectrum
which was flux-calibrated using 2MASS photometry.\\
References: (1) Warren et al. 2007, (2) Delorme et al. 2008a, 
(3) Tinney et al. 2005, (4) Burgasser et al. 2008,
(5) Burningham et  al. 2008.\\
\end{table}


\clearpage

\begin{table}
\footnotesize
\caption{\label{tab2}Likely Range of Properties of the Sample.}
\begin{tabular}{@{}lcccccccc}
\br
Name & $T_{\rm eff}$  & $g$  & [m/H] & R/R$_{\odot}$ & Mass  & Age & Distance  & V$_{\rm tan}$  \\
     & (K) &   (ms$^{-2}$) &  &  &   (Jupiter) &  (Gyr) & (pc) & (kms$^{-1}$) \\
\mr
ULAS J0034$-$00 & 550 & 100 & 0.0 & 0.12 & 5 -- 8 & 0.1 -- 0.2 & 14  & 25 \\ 
          & 550 & 300 & $+$0.3 & 0.11 & 10 -- 20 & 1 -- 2     & 13  & 23 \\
          & 600 & 100 & 0.0    & 0.12 &  5 -- 8 & 0.1 -- 0.2  & 16  & 28 \\
          & 600 & 300 & 0.0    & 0.11 & 10 -- 20 & 1 -- 2     & 13  & 25 \\
ULAS J1335$+$11 & 500 & 100 & 0.0 & 0.12 &  5 -- 8 & 0.1 -- 0.2  & 9  & 17 \\
          & 500 & 300 & $+$0.3 & 0.11 & 10 -- 20 & 1 -- 2     &  9   & 14 \\
          & 550 & 100 & 0.0    & 0.12 &  5 -- 8 & 0.1 -- 0.2  & 11  & 17 \\
2MASS J0939$-$24AB$^{\rm a}$ & 600 & 1000 & $-0.3$ -- 0.0 & 0.09 & 20 -- 40 & 2 -- 10 & 5.34 & 30 \\
 2MASS J0939$-$24A$^{\rm b}$ & 700 & 1000 -- 2000 & $-0.3$ -- 0.0 & 0.08 -- 0.09 & 30 -- 50 & 4 -- 10 & 5.34 & 30 \\
 2MASS J0939$-$24B$^{\rm b}$ & 500 & 1000 & $-0.3$ -- 0.0 & 0.09 & 20 -- 30 & 4 -- 10 & 5.34 & 30 \\
CFBD 0059$-$01 & 550 & 300 & 0.0 & 0.11 & 10 --20 & 0.5 -- 2  & 11 & 50 \\
                & 550 & 1000 & $+$0.3 & 0.09 & 25 -- 40 & 4 -- 10 & 11 & 50 \\
                & 550 & 2000 & $+$0.3 & 0.08 & 40 -- 50 &   10    & 10 & 45 \\
                & 600 & 300 & 0.0 &  0.11 & 10 --20 & 0.5 -- 2  & 14 & 64 \\
                & 600 & 1000 & 0.0 & 0.09 & 25 -- 40 & 4 -- 10  & 11 & 50\\
                & 600 & 2000 & 0.0 & 0.08 & 40 -- 50 &   10    & 10 & 45 \\
\br
\end{tabular}
\noindent
$^{\rm a}$Assuming that this object is a pair of similar luminosity dwarfs.\\
$^{\rm b}$Assuming that this object is a pair of dissimilar  dwarfs.\\
\end{table}

\clearpage





\begin{figure}
\includegraphics[height=.6\textheight,angle=-90]{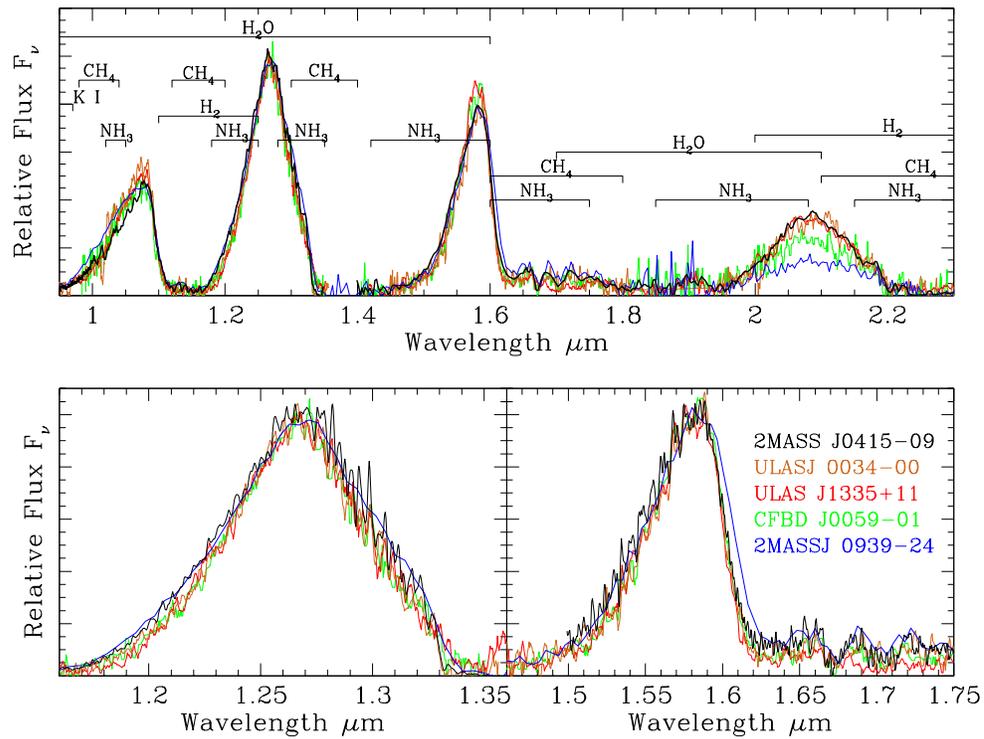}
  \caption{Near-infrared spectra of the three T9 dwarfs ULAS J0034$-$00, ULAS J1335$+$11 and CFBD J0059$-$01,  and the T8 dwarfs 2MASS J0415$-$09 and 2MASS J0939$-$24. The top panel shows the spectra normalized to the 1.25 $\mu m$ flux peak; the lower panels show the $J$- and $H$-band  regions with the spectra scaled to the 1.25 $\mu m$ and 1.6 $\mu m$ peaks. Data sources are: Burgasser et al. 2008, Burningham et al. 2008,
Delorme et al. 2008a, Knapp et al. 2004, McLean et al. 2003, Warren et al. 2007. The principal absorbing species are 
indicated in the top panel.}
\end{figure}

\clearpage \begin{figure} \includegraphics[height=.6\textheight,angle=-90]{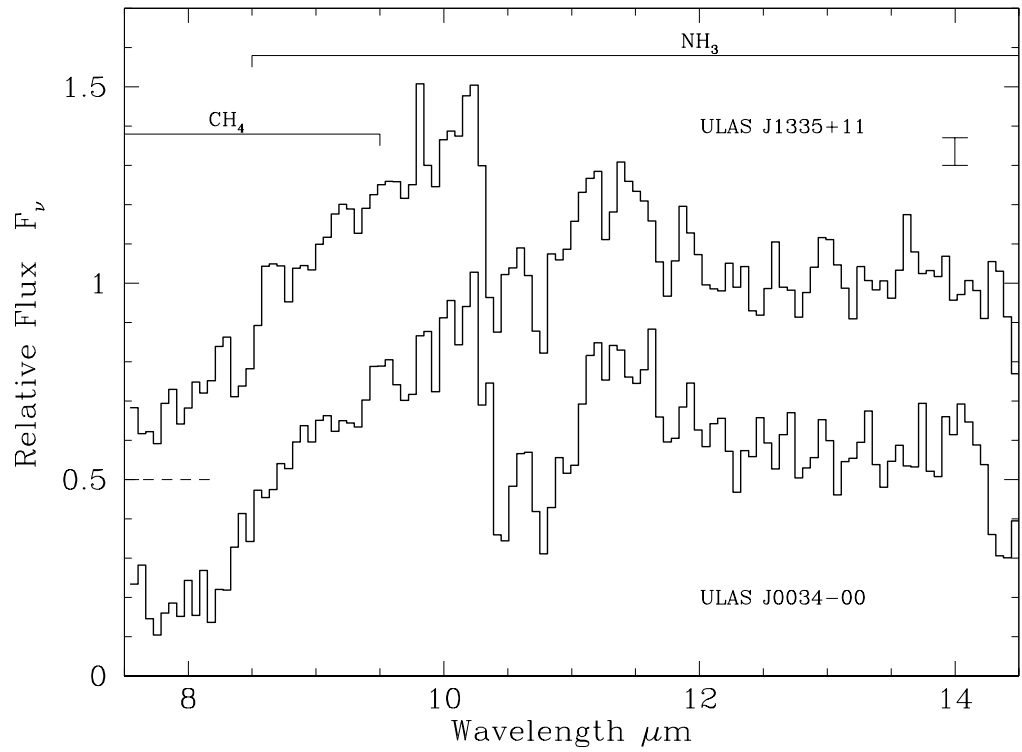} \caption{{\it Spitzer} IRS 
spectra of ULAS J0034$-$00 and ULAS J1335$+$11, normalized to the flux peak at 10.2 $\mu$m and offset for clarity; 
a typical error bar is shown. Dashed lines show the zero flux level for ULAS J1335$+$11. The 
principal absorbing species are indicated. } \end{figure}

\clearpage \begin{figure} 
\includegraphics[height=.8\textheight]{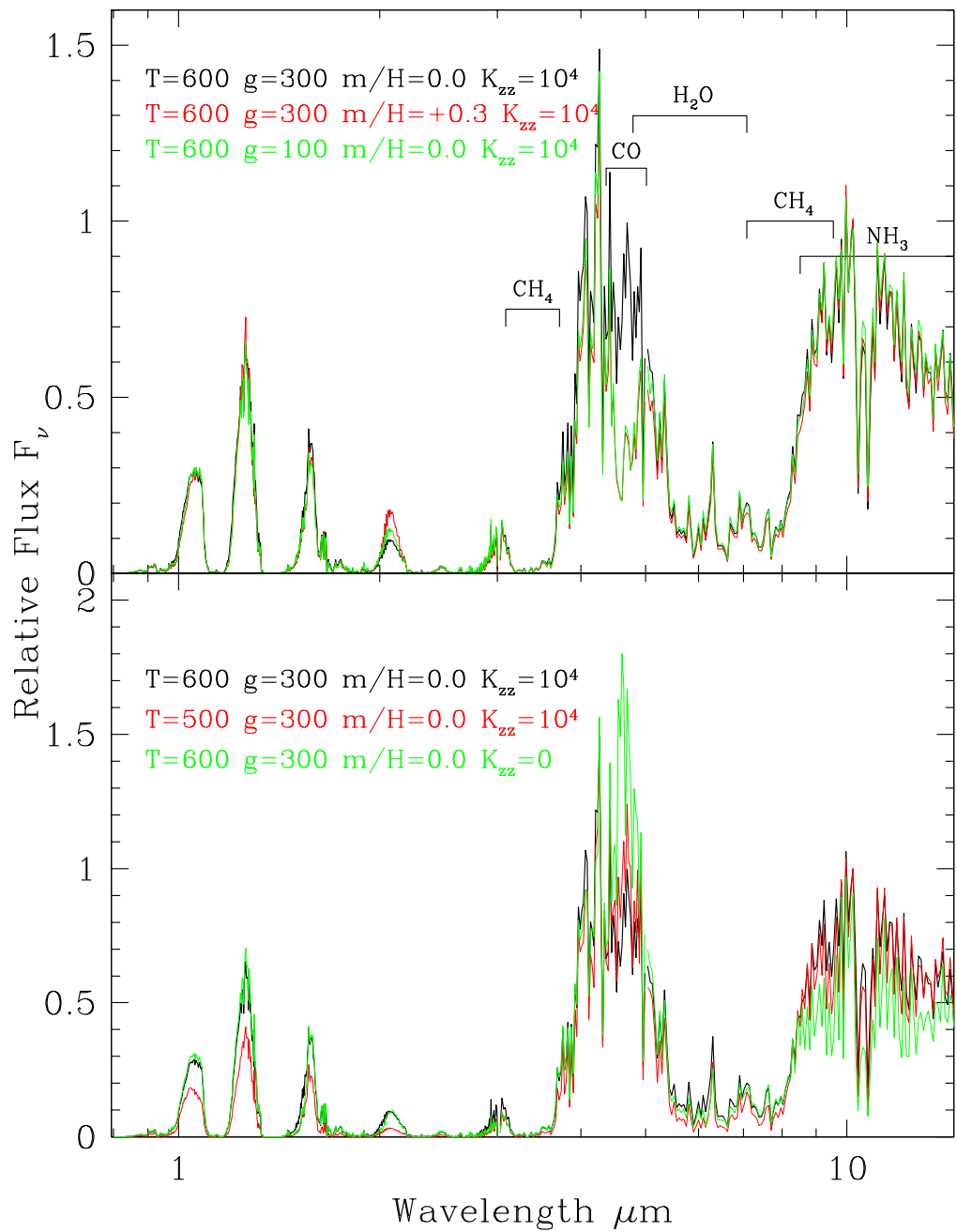} \caption{Synthetic 
spectra demonstrating the effect of varying the atmospheric 
parameters; the spectra are normalized to unity at $\lambda =$ 10 $\mu$m. In 
both panels black curves are the baseline model with $T_{\rm eff}=$ 600~K, 
$g =$ 300 ms$^{-2}$, [m/H]$=$0 and $K_{zz}=10^4$ cm$^{2}$ s$^{-1}$. The 
top panel demonstrates the similar effects of increasing metallicity and 
decreasing gravity -- the $2.2 \mu$m flux is increased while the $4.5 
\mu$m flux is decreased. The lower panel demonstrates that reducing 
$T_{\rm eff}$ reddens the slope of the energy distribution, and removing 
chemical mixing increases the $4.5 \mu$m flux while decreasing the 9 -- 15 
$\mu$m flux. The principal absorbing species in the mid-infrared are indicated in the top panel.} \end{figure}

\clearpage
\begin{figure}
\includegraphics[height=.8\textheight]{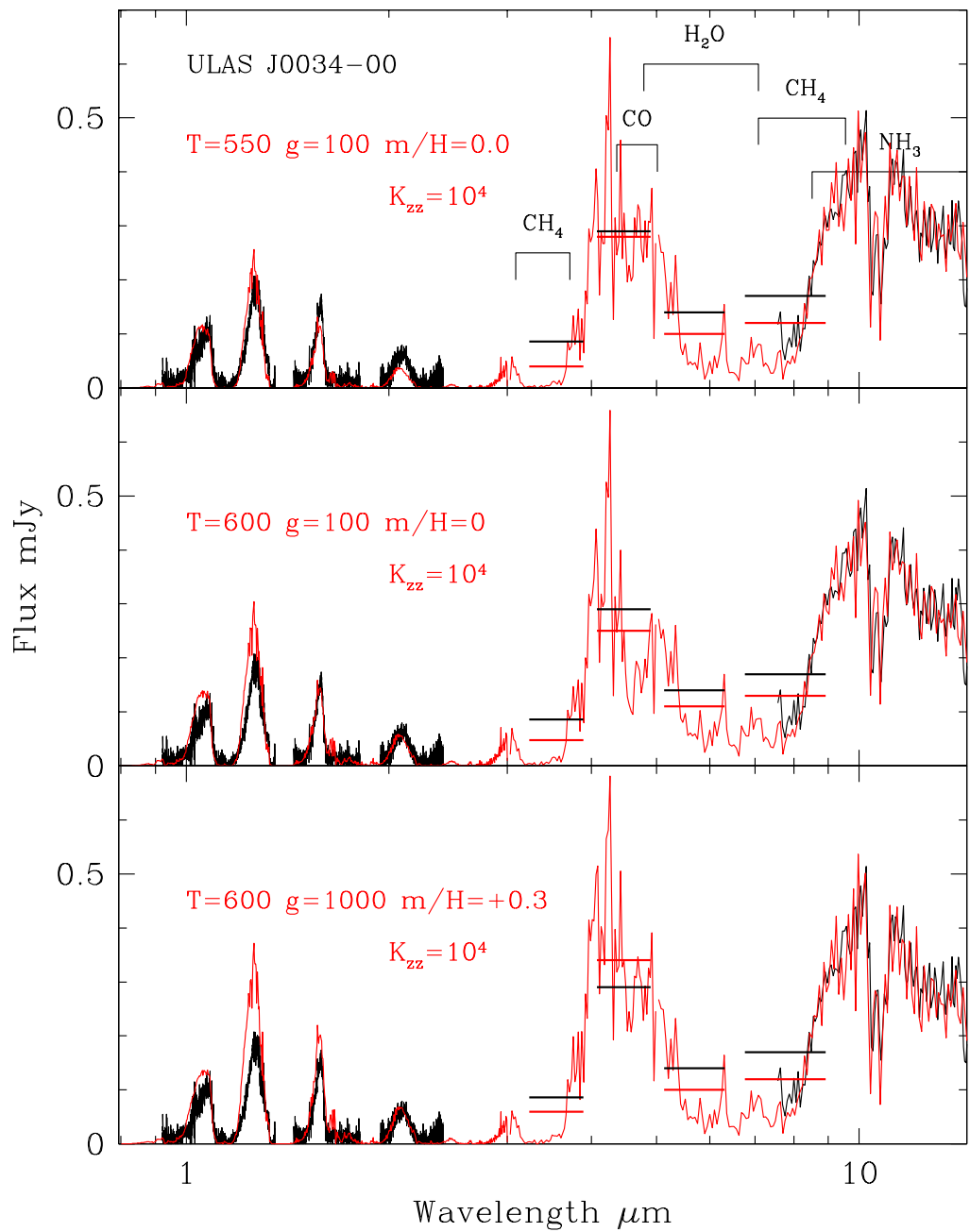}
\caption{Near- and mid-infrared spectra of ULAS J0034$-$00 (black lines)
compared to various synthetic spectra (red lines) generated by models with parameters
as indicated in the legends. Scaling of the models to the data is done by minimising
the weighted least-squares fit; this provides a distance to the dwarf.
Horizontal lines show observed and calculated IRAC fluxes at this distance. 
The uncertainty in the measured IRAC fluxes is 3 -- 6 \%.
A gravity as high as $g=1000$  ms$^{-2}$ is excluded by the poorer spectral fit
shown in the bottom panel.
}
\end{figure}

\clearpage
\begin{figure}
\includegraphics[height=.8\textheight]{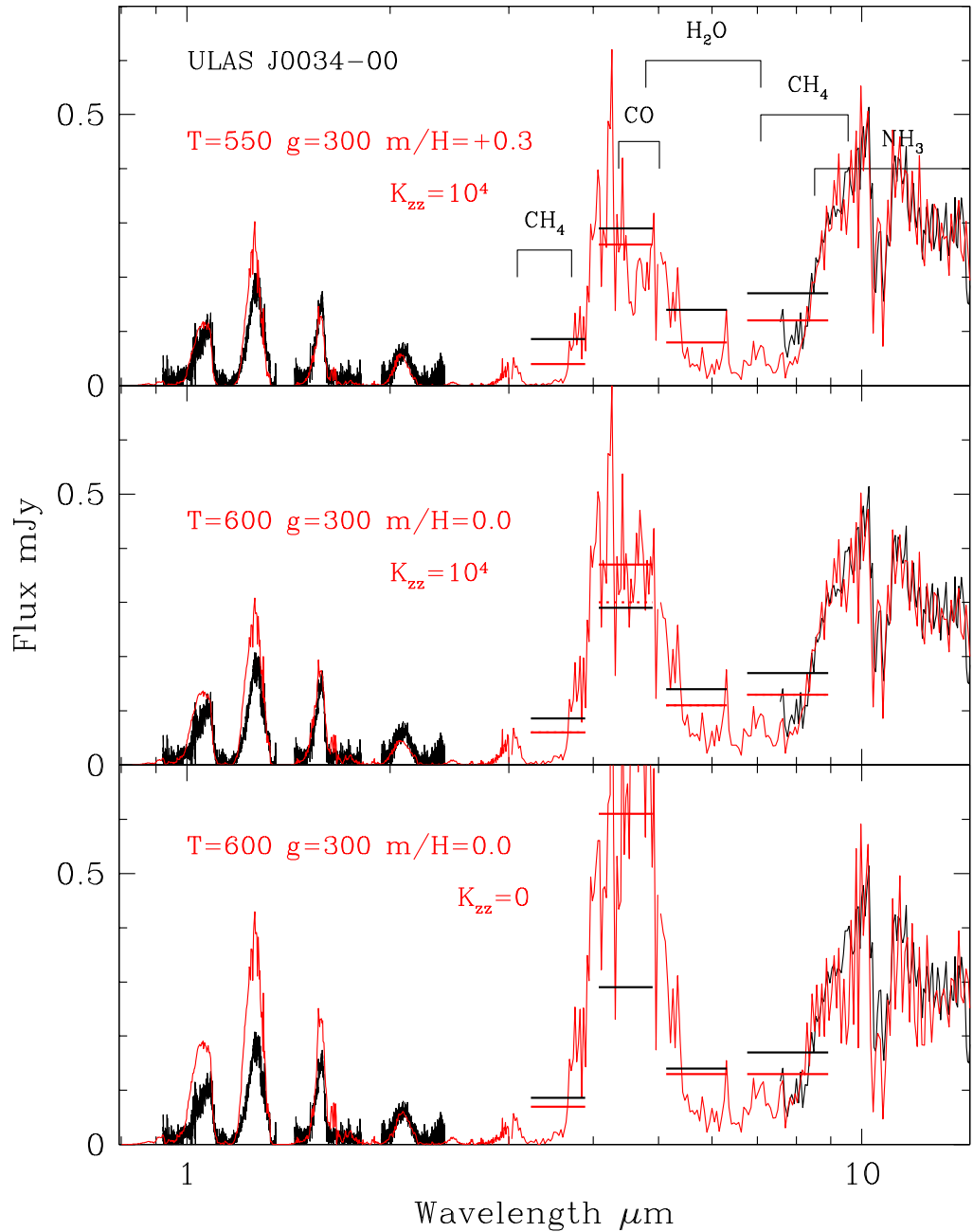}
\caption{Near- and mid-infrared spectra of ULAS J0034$-$00 (black lines)
compared to various synthetic spectra and photometry (red lines) as in Figure 4.
Dotted red lines in the middle panel shows the effect 
on the IRAC fluxes of increasing the mixing coefficient to  $K_{zz}=10^6$ cm$^{2}$ s$^{-1}$.
Equilibrium chemistry,  $K_{zz}=0$ cm$^{2}$ s$^{-1}$, models can be excluded;
the bottom panel shows that in this case reproducing the mid-infrared spectrum leads to
too high near-infrared flux levels, and also a discrepancy of a factor of $\sim2$
at 4.5 $\mu$m (see also the lower panel of Figure 3). 
}
\end{figure}

\clearpage
\begin{figure}
\includegraphics[height=.8\textheight]{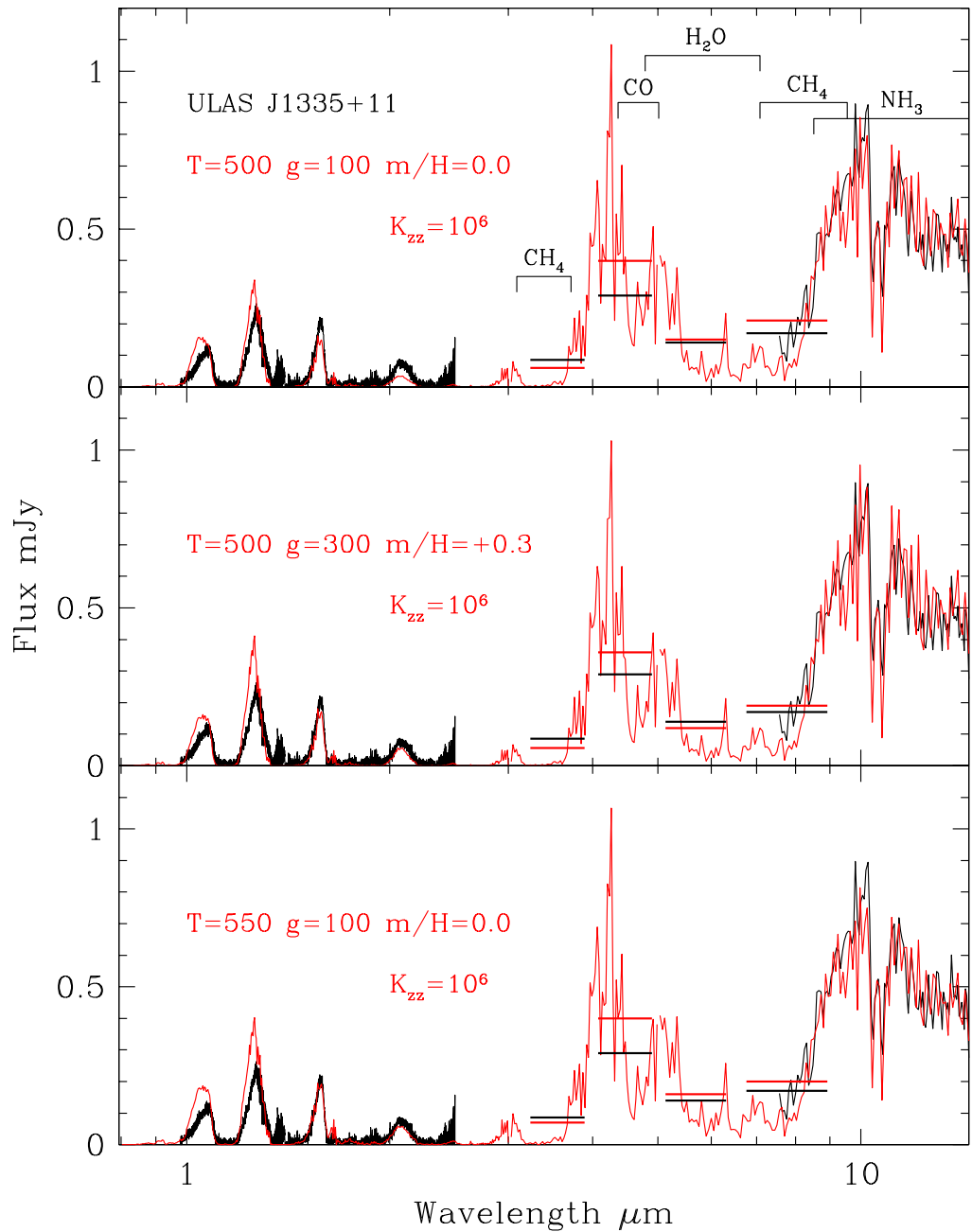}
\caption{Fits (red lines) to ULAS J1335$+$11 (black lines) using models with parameters
as indicated in the legends, as in Figure 4.
The uncertainty in the measured IRAC fluxes is 3 -- 7 \%.
 }
\end{figure}

\clearpage
\begin{figure}
\includegraphics[height=.8\textheight,]{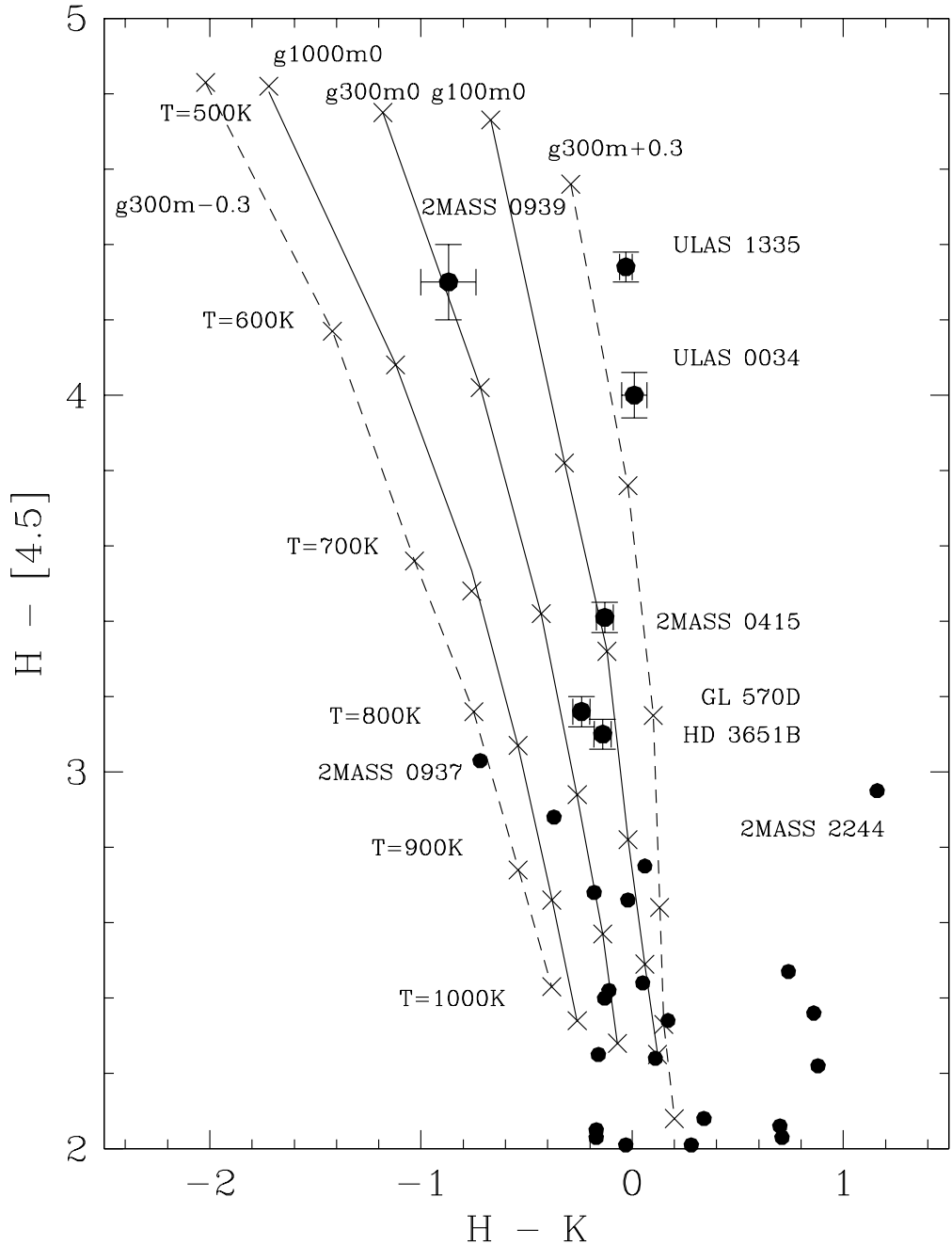}
\caption{Calculated sequences with $T_{\rm eff}$ in the colors $H - K$:$H - $[4.5] 
are shown as solid and dashed lines. Solid lines are solar metallicity models with gravities
1000, 300 and 100 ms$^{-2}$ from left to right. Dashed lines have $g = 300$ ms$^{-2}$
and demonstrate the effect of varying metallicity: [m/H] $= -0.3$ on the left and 
[m/H] $=+0.3$ on the right. Crosses mark $T_{\rm eff}=$ 500, 600, 700, 800, 900 and 1000 K from top to bottom. 
All sequences are for non-equilibrium chemistry with $K_{zz}=10^4$ cm$^{2}$ s$^{-1}$.
At these temperatures the range of gravities shown corresponds to an age range
of $\sim$ 0.1 -- 5 Gyr (right to left).
Solid points are observed colors for dwarfs discussed in the text and others
taken from the literature.  See \S 5.3.
}
\end{figure}

\clearpage
\begin{figure}
\includegraphics[height=.8\textheight]{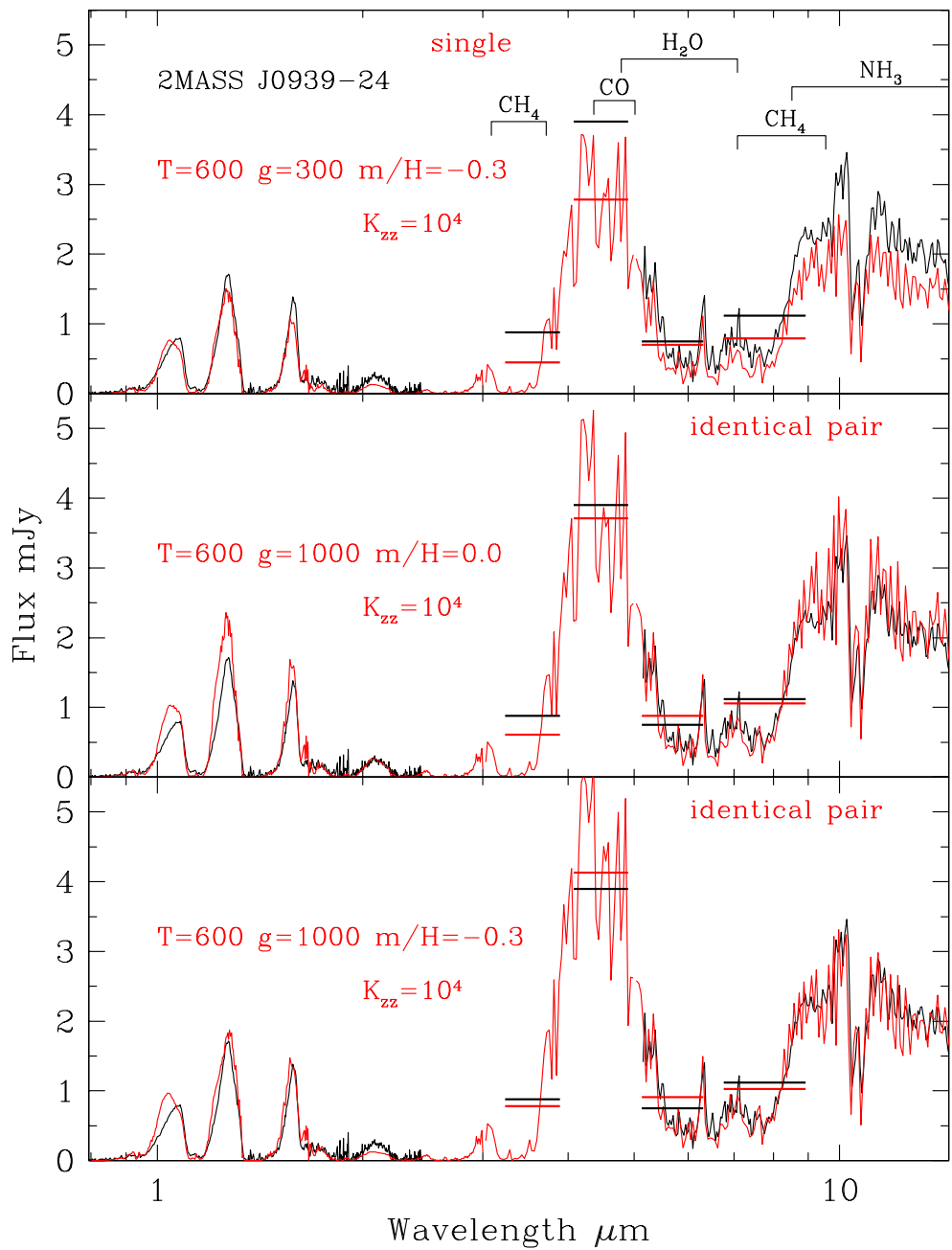}
\caption{Fits (red lines) to 2MASS J0939$-$24 (black lines) using models with parameters
as indicated in the legends. The top panel assumes the dwarf is a single object, the lower panels that it is an 
identical pair of objects. Horizontal lines show observed and calculated IRAC fluxes.  
The uncertainty in the measured IRAC fluxes is 4 \%. }
\end{figure}

\clearpage
\begin{figure}
\includegraphics[height=.8\textheight]{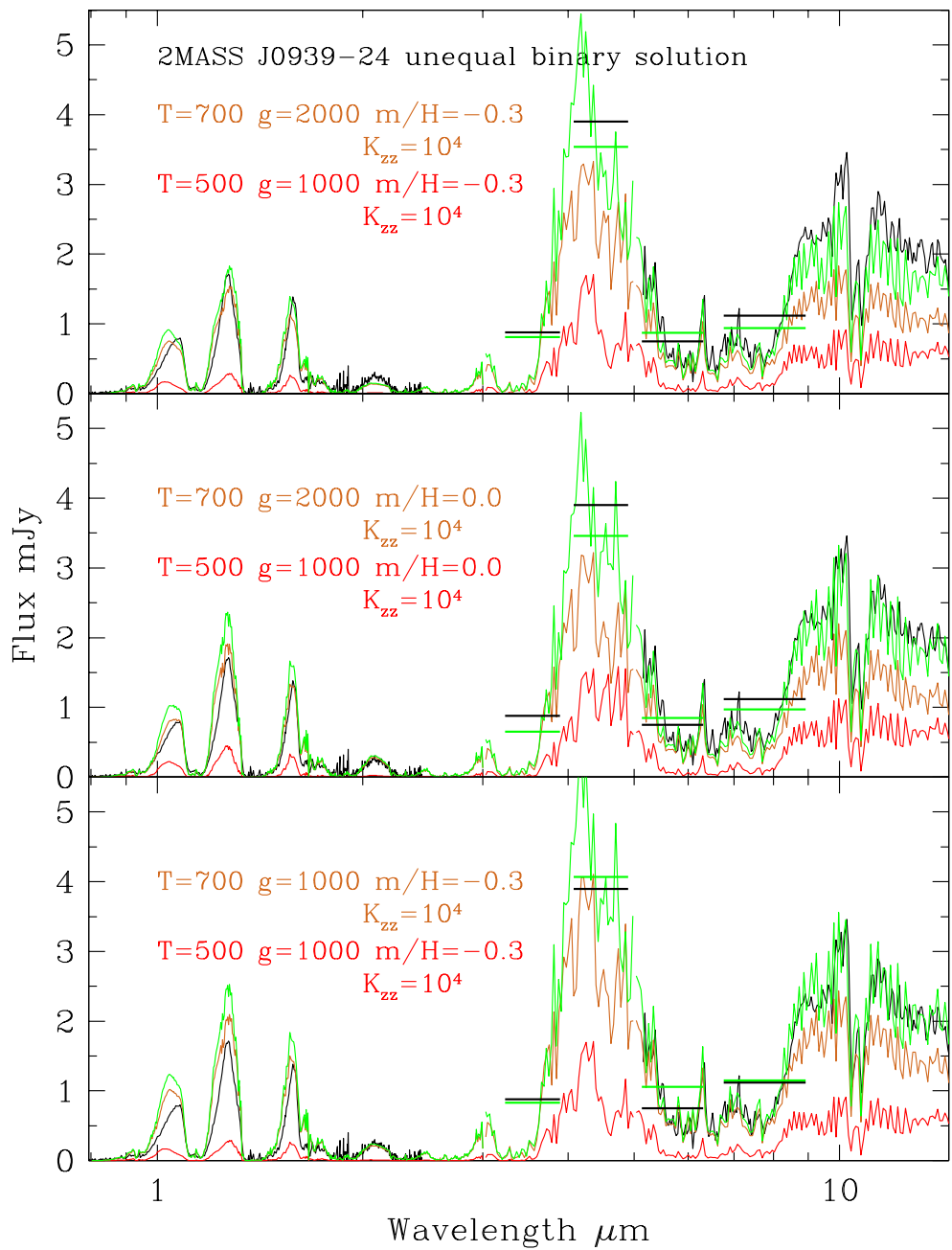}
\caption{Fits to 2MASS J0939$-$24 (black line) as a non-equal luminosity binary. The 500 K
component is shown in red, the 700 K in brown and the combined spectrum in green.
Horizontal lines show observed and calculated IRAC fluxes where the IRAC fluxes are synthesized
from the combined model spectrum. The uncertainty in the measured IRAC fluxes is 4 \%.
 }
\end{figure}

\end{document}